\documentclass[%
reprint,
showpacs,preprintnumbers,
nofootinbib,
amsmath,amssymb,
aps,
prd,
floatfix,
]{revtex4-1}

\usepackage{graphicx}
\usepackage{dcolumn}
\usepackage{bm}

\usepackage{url}

\newcommand{\boss}[2]{\ensuremath{\rlap{\kern-2.5pt\ensuremath{\overset{\scriptscriptstyle(-)}{\phantom{#1}}}}{\ensuremath{{#1}_{#2}}}}}

\begin{document}

\preprint{\begin{tabular}{l}
\texttt{EURONU-WP6-11-44}
\\
\texttt{arXiv:1111.1069 [hep-ph]}
\end{tabular}}

\title{Implications of 3+1 Short-Baseline Neutrino Oscillations}

\author{Carlo Giunti}
\email{giunti@to.infn.it}
\altaffiliation[also at ]{Department of Theoretical Physics, University of Torino, Italy}
\affiliation{INFN, Sezione di Torino, Via P. Giuria 1, I--10125 Torino, Italy}

\author{Marco Laveder}
\email{laveder@pd.infn.it}
\affiliation{Dipartimento di Fisica ``G. Galilei'', Universit\`a di Padova,
and
INFN, Sezione di Padova,
Via F. Marzolo 8, I--35131 Padova, Italy}

\date{\today}

\begin{abstract}
We present an upgrade of the
3+1 global fit of short-baseline neutrino oscillation data
obtained with the addition of
KARMEN
and
LSND
$\nu_{e} + {}^{12}\text{C} \to {}^{12}\text{N}_{\text{g.s.}} + e^{-}$
scattering data.
We discuss the implications for the measurements of the effective neutrino mass
in $\beta$-decay and neutrinoless double-$\beta$-decay
experiments.
We find respective predicted ranges of about 0.1--0.7 eV
and
0.01--0.1 eV.
\end{abstract}


\maketitle

\section{Introduction}
\label{Introduction}

The study of active-sterile neutrino oscillations
received recently a boost from indications in favor of possible
short-baseline oscillations of two types:
1) $\bar\nu_{\mu}\to\bar\nu_{e}$ oscillations
observed in the LSND
\cite{hep-ex/0104049}
and
MiniBooNE
\cite{1007.1150}
experiments;
2) $\bar\nu_{e}$ and $\nu_{e}$ disappearance
revealed, respectively, by the
Reactor Anomaly
\cite{1101.2755}
and the
Gallium Anomaly
\cite{Anselmann:1995ar,Hampel:1998fc,1001.2731,Abdurashitov:1996dp,hep-ph/9803418,nucl-ex/0512041,0901.2200}.
In this paper we consider these indications in the framework of
hierarchical 3+1 neutrino mixing
and we discuss the implications for the measurements of the effective neutrino mass
in $\beta$-decay and neutrinoless double-$\beta$-decay
experiments.
We also upgrade the
global fit presented in Ref.~\cite{1109.4033}
by the addition of
KARMEN \cite{Bodmann:1994py,hep-ex/9801007}
and
LSND \cite{hep-ex/0105068}
$\nu_{e} + {}^{12}\text{C} \to {}^{12}\text{N}_{\text{g.s.}} + e^{-}$
scattering data, as suggested in Ref.~\cite{1106.5552}.

Short-baseline (SBL) neutrino oscillations are generated by a squared-mass difference
$\Delta{m}^2_{\text{SBL}} \gtrsim 0.1 \, \text{eV}^2$,
which is much larger than
the two measured solar (SOL) and atmospheric (ATM) squared-mass differences
$
\Delta{m}^2_{\text{SOL}}
=
(7.6 \pm 0.2) \times 10^{-5} \, \text{eV}^2
$
\cite{1010.0118}
and
$
\Delta{m}^2_{\text{ATM}}
=
2.32 {}^{+0.12}_{-0.08} \times 10^{-3} \, \text{eV}^2
$
\cite{1103.0340}.
The minimal neutrino mixing schemes which can provide a third squared-mass difference
for short-baseline neutrino oscillations
require the introduction of a sterile neutrino $\nu_{s}$
(see Refs.~\cite{hep-ph/9812360,hep-ph/0405172,hep-ph/0606054,GonzalezGarcia:2007ib}).
Hierarchical
3+1 neutrino mixing is a perturbation of the standard three-neutrino mixing in which
the three active neutrinos
$\nu_{e}$,
$\nu_{\mu}$,
$\nu_{\tau}$
are mainly composed of three massive neutrinos
$\nu_1$,
$\nu_2$,
$\nu_3$
with light masses
$m_1$,
$m_2$,
$m_3$,
such that
$
\Delta{m}^2_{\text{SOL}}
=
\Delta{m}^2_{21}
$
and
$
\Delta{m}^2_{\text{ATM}}
=
|\Delta{m}^2_{31}|
\simeq
|\Delta{m}^2_{32}|
$,
with the standard notation
$\Delta{m}^2_{kj} \equiv m_{k}^2 - m_{j}^2$
(see Ref.~\cite{Giunti-Kim-2007}).
The sterile neutrino is mainly composed of a heavy neutrino
$\nu_{4}$
with mass $m_{4}$
such that
$\Delta{m}^2_{\text{SBL}} = \Delta{m}^2_{41}$
and
\begin{equation}
m_{1}
\,,\,
m_{2}
\,,\,
m_{3}
\ll
m_{4}
\quad
\Rightarrow
\quad
m_{4} \simeq \sqrt{\Delta{m}^2_{41}}
\,.
\label{hierarchy}
\end{equation}
Under these hypotheses,
the effects of active-sterile neutrino mixing
in solar
\cite{0910.5856,1105.1705}
and atmospheric
\cite{0709.1937,1104.1390,1108.4360,1109.5748}
neutrino experiments are small,
but should be revealed sooner or later.

In 3+1 neutrino mixing,
the effective flavor transition and survival probabilities
in short-baseline neutrino oscillation experiments
are given by
(see Refs.~\cite{hep-ph/9812360,hep-ph/0405172,hep-ph/0606054,GonzalezGarcia:2007ib})
\begin{align}
\null & \null
P_{\boss{\nu}{\alpha}\to\boss{\nu}{\beta}}^{\text{SBL}}
=
\sin^{2} 2\vartheta_{\alpha\beta}
\sin^{2}\left( \frac{\Delta{m}^2_{41} L}{4E} \right)
\qquad
(\alpha\neq\beta)
\,,
\label{trans}
\\
\null & \null
P_{\boss{\nu}{\alpha}\to\boss{\nu}{\alpha}}^{\text{SBL}}
=
1
-
\sin^{2} 2\vartheta_{\alpha\alpha}
\sin^{2}\left( \frac{\Delta{m}^2_{41} L}{4E} \right)
\,,
\label{survi}
\end{align}
for
$\alpha,\beta=e,\mu,\tau,s$,
with the transition amplitudes
\begin{align}
\null & \null
\sin^{2} 2\vartheta_{\alpha\beta}
=
4 |U_{\alpha4}|^2 |U_{\beta4}|^2
\,,
\label{transsin}
\\
\null & \null
\sin^{2} 2\vartheta_{\alpha\alpha}
=
4 |U_{\alpha4}|^2 \left( 1 - |U_{\alpha4}|^2 \right)
\,.
\label{survisin}
\end{align}

The hierarchical 3+1 scheme
may be compatible with the results of standard cosmological $\Lambda$CDM
analyses of the
Cosmic Microwave Background and Large-Scale Structures data,
which constrain the three light neutrino masses
to be much smaller than 1 eV
\cite{0805.2517,0910.0008,0911.5291,1006.3795}
and
indicate that 
one or two sterile neutrinos may have been thermalized in the early Universe
\cite{1006.5276,1102.4774,1104.0704,1104.2333,1106.5052,1108.4136,1109.2767,1110.4271},
with a upper limit of the order of 1 eV on their masses.
Also
Big-Bang Nucleosynthesis data
\cite{astro-ph/0408033,1001.4440}
are compatible with the existence of sterile neutrinos,
with the indication however that the thermalization of
more than one sterile neutrino is disfavored
\cite{1103.1261,1108.4136}.

\begin{figure}[t!]
\begin{center}
\includegraphics*[width=\linewidth]{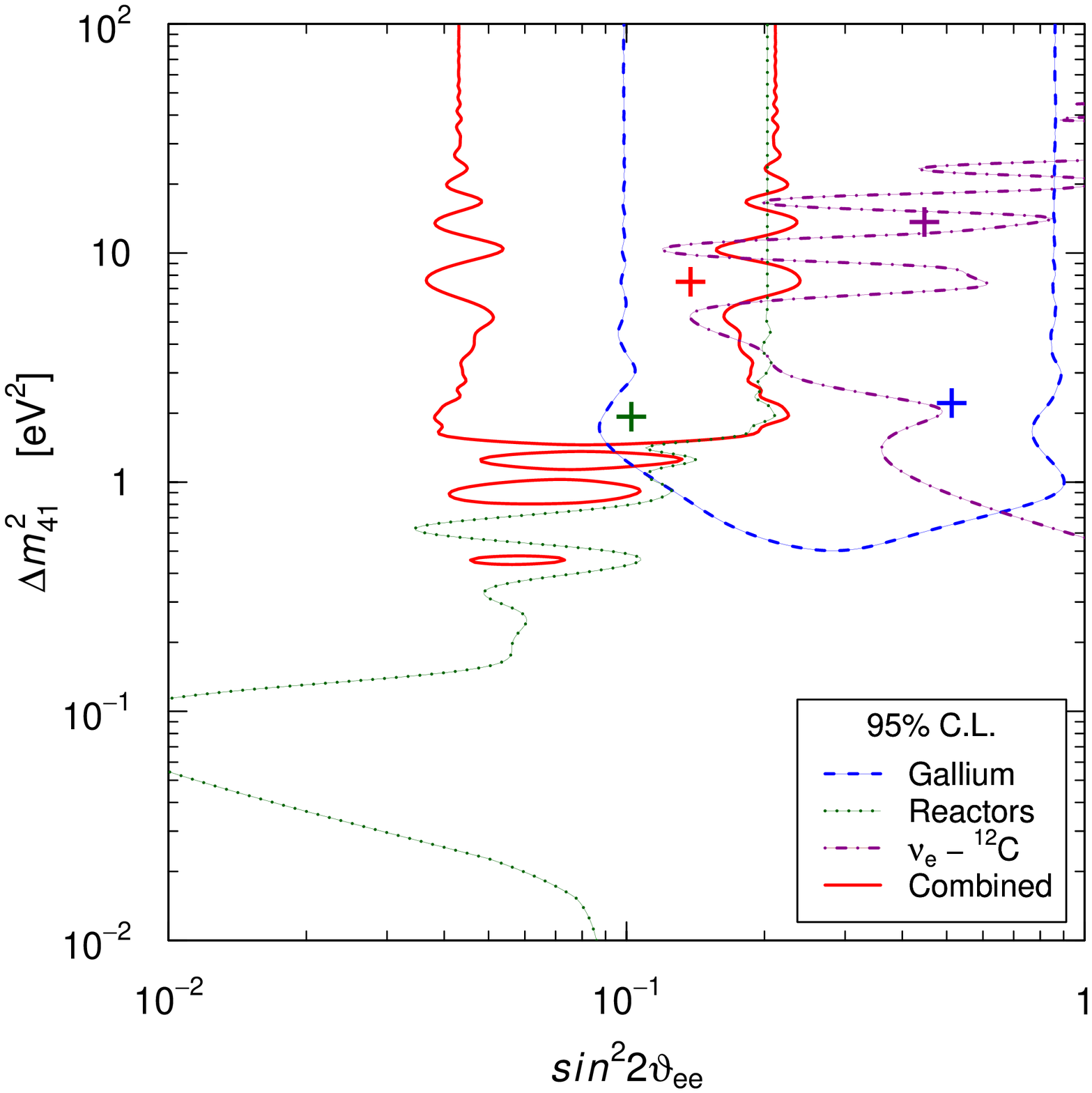}
\end{center}
\caption{ \label{sup-gal-rea-nec}
Superposition of the 95\% C.L. contours
in the
$\sin^{2}2\vartheta_{ee}$--$\Delta{m}^{2}_{41}$ plane
obtained from the separate fits of Gallium, reactor and $\nu_{e}-{}^{12}\text{C}$ data
and that obtained from the combined fit.
The best-fit points are indicated by crosses
(see Table.~\ref{tab-bef}).
}
\end{figure}

\begin{figure}[t!]
\begin{center}
\includegraphics*[width=\linewidth]{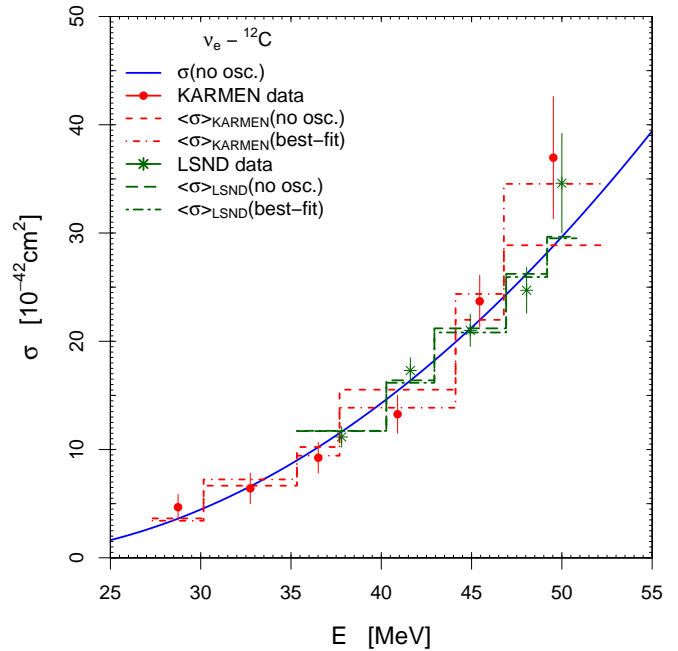}
\end{center}
\caption{ \label{hst-nec}
KARMEN and LSND $\nu_{e}-{}^{12}\text{C}$ cross section data (points and asterisks, respectively)
with the corresponding statistical error bars.
The solid blue line shows the best-fit dependence of the cross section on the neutrino energy $E$
in the case of no oscillations.
The dashed red and long-dashed green histograms show, respectively, the corresponding average cross section for the energy bins of the
KARMEN and LSND data.
The dash-dotted red and long-dash-dotted green histograms show, respectively, the average cross section modulated
by the best-fit oscillation probability for the energy bins of the
KARMEN and LSND data.
}
\end{figure}

\begin{figure*}[t!]
\begin{center}
\begin{tabular}{c}
\includegraphics*[width=0.8\linewidth]{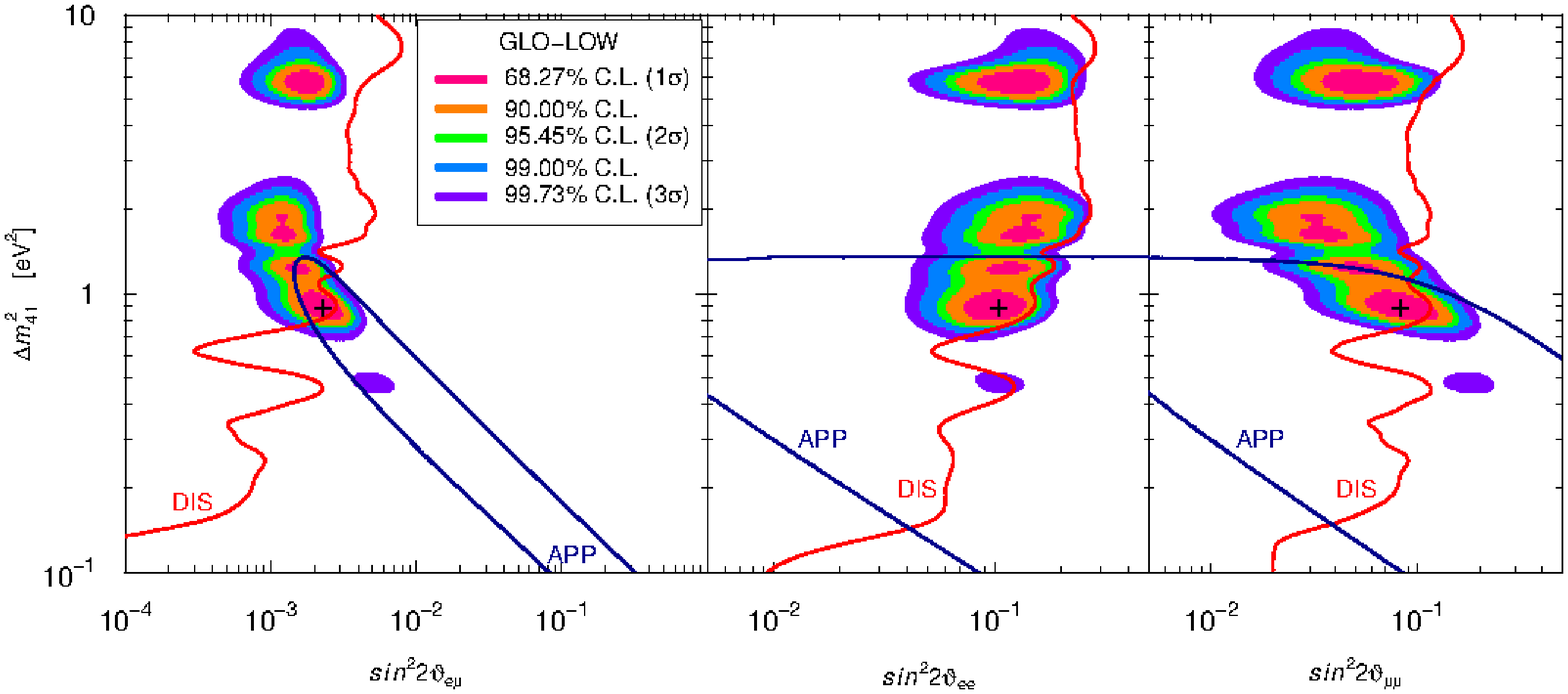}
\\
\includegraphics*[width=0.8\linewidth]{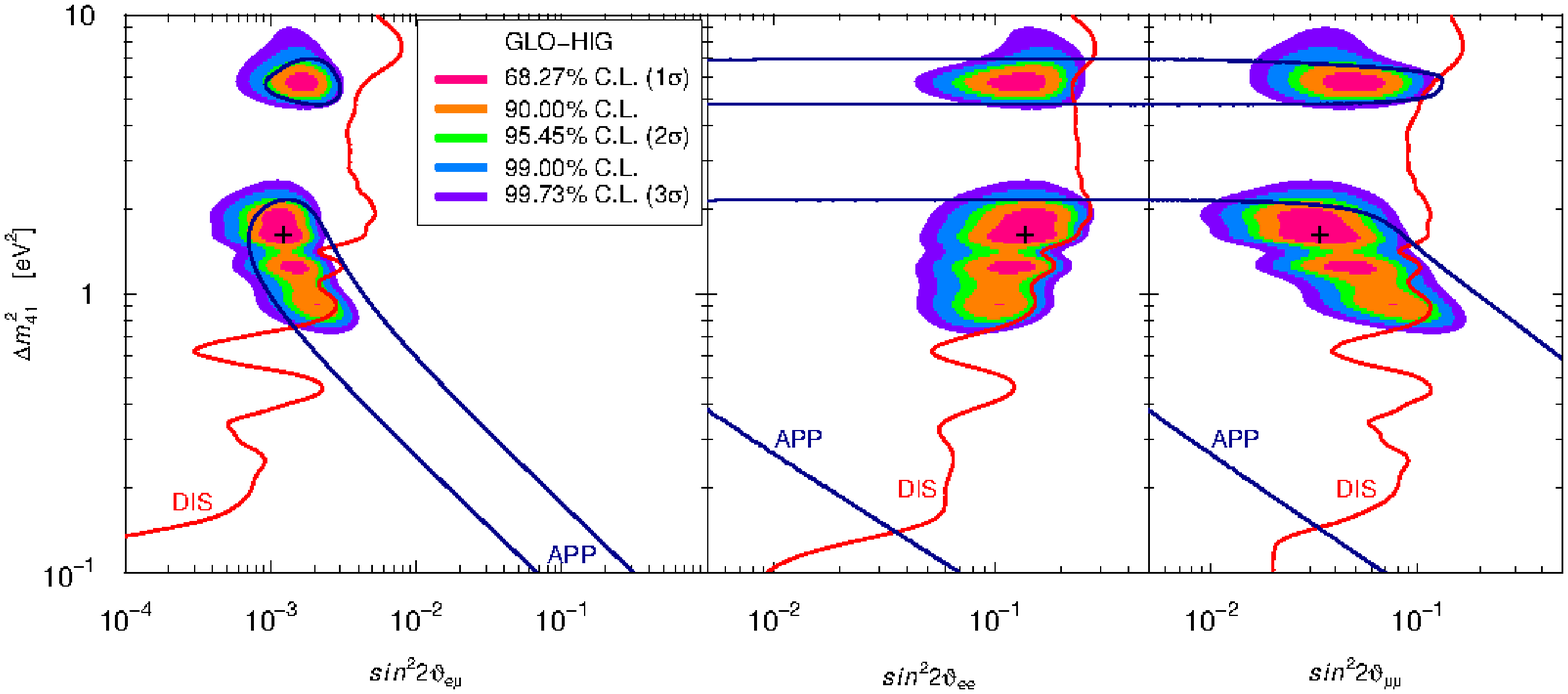}
\end{tabular}
\end{center}
\caption{ \label{con-glo}
Allowed regions in the
$\sin^{2}2\vartheta_{e\mu}$--$\Delta{m}^2_{41}$,
$\sin^{2}2\vartheta_{ee}$--$\Delta{m}^2_{41}$ and
$\sin^{2}2\vartheta_{\mu\mu}$--$\Delta{m}^2_{41}$
planes obtained from the GLO-LOW and GLO-HIG
global analyses of short-baseline neutrino oscillation data
(see Tab~\ref{tab-bef}).
The best-fit points are indicated by crosses (see Table.~\ref{tab-bef}).
The thick solid blue lines with the label APP
show the $3\sigma$ allowed regions obtained from the analysis of
$\protect\boss{\nu}{\mu}\to\protect\boss{\nu}{e}$
appearance data.
The thick solid red lines with the label DIS
show the $3\sigma$ allowed regions obtained from the analysis of
disappearance data.
}
\end{figure*}

\begin{figure}[t!]
\begin{center}
\includegraphics*[width=\linewidth]{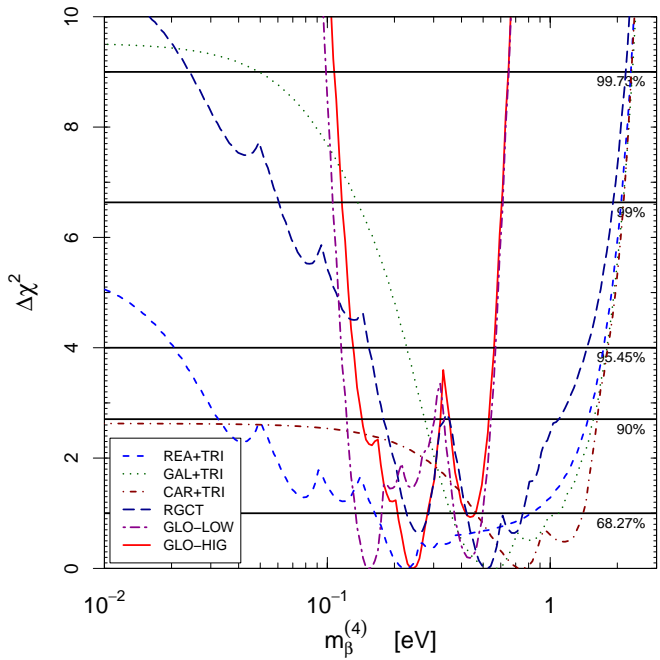}
\end{center}
\caption{ \label{mbt-chi-plt}
Marginal
$\Delta\chi^2 = \chi^2 - \chi^2_{\text{min}}$
as a function of the contribution
$m_{\beta}^{(4)} = |U_{e4}| \sqrt{\Delta{m}^2_{41}}$
to the effective $\beta$-decay electron-neutrino mass $m_{\beta}$
obtained from
the fits of Tritium (TRI) data with
Reactor (REA),
Gallium (GAL) and
$\nu_{e}$-${}^{12}\text{C}$ (CAR)
data,
their combined fit (RGCT) and
the GLO-LOW and GLO-HIG global fits.
}
\end{figure}

\begin{figure}[t!]
\begin{center}
\includegraphics*[width=\linewidth]{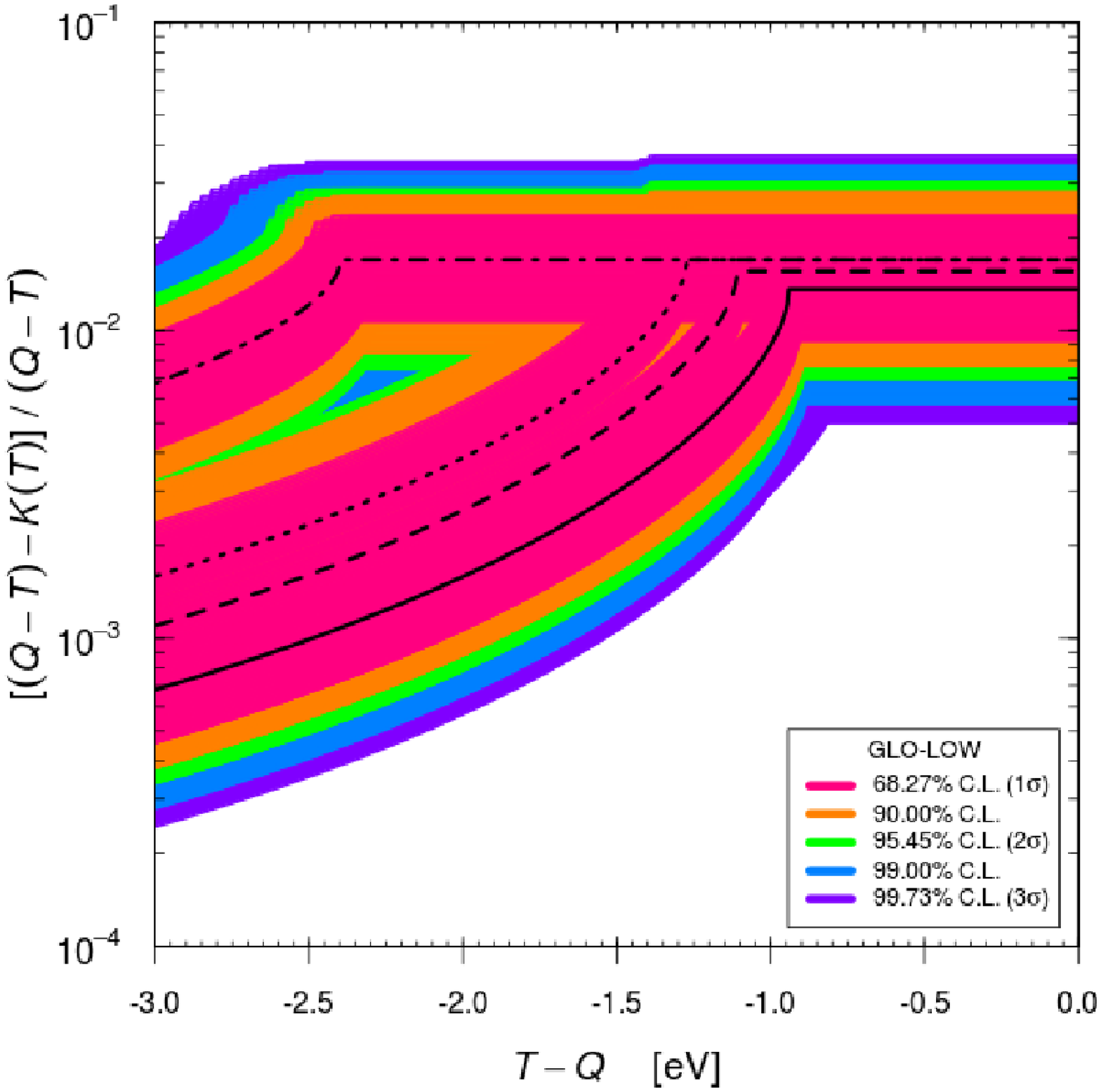}
\end{center}
\caption{ \label{plt-kur-dif-low}
Bands of the relative deviation of the Kurie plot in $\beta$-decay
corresponding to the allowed regions in the
$\sin^{2}2\vartheta_{ee}$--$\Delta{m}^2_{41}$
plane in Fig.~\ref{con-glo},
obtained from the GLO-LOW global analysis of short-baseline neutrino oscillation data
(see Tab~\ref{tab-bef}).
The black solid line corresponds to the best-fit point
($m_{4} = \protect0.94 \, \text{eV}$
and
$|U_{e4}|^2 = \protect0.027$).
The dashed, dotted and dash-dotted lines correspond,
respectively,
to the local minima at
($m_{4} = \protect1.11 \, \text{eV}$,
$|U_{e4}|^2 = \protect0.03$),
($m_{4} = \protect1.27 \, \text{eV}$,
$|U_{e4}|^2 = \protect0.035$) and
($m_{4} = \protect2.40 \, \text{eV}$,
$|U_{e4}|^2 = \protect0.033$).
}
\end{figure}

\begin{figure}[t!]
\begin{center}
\includegraphics*[width=\linewidth]{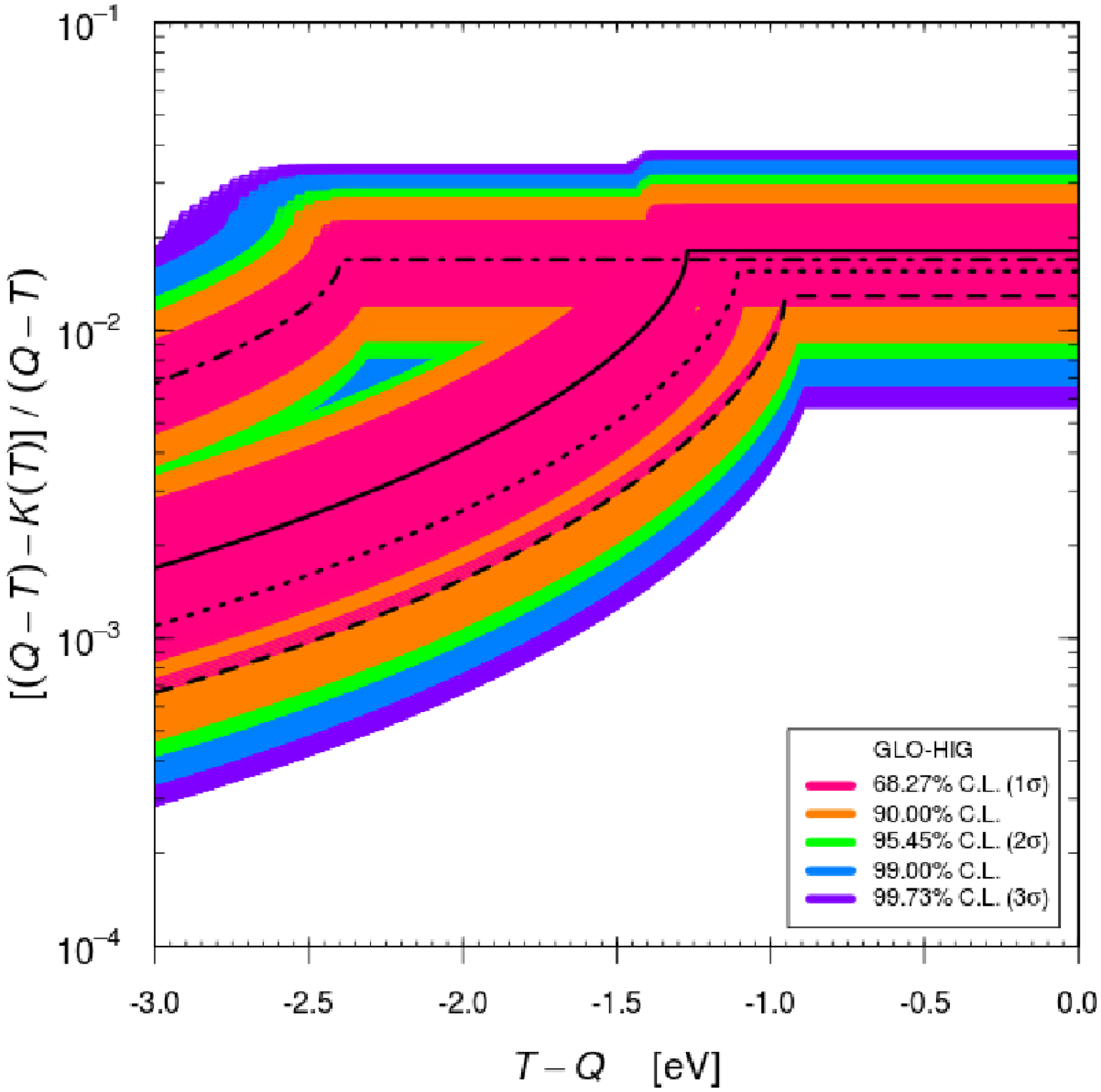}
\end{center}
\caption{ \label{plt-kur-dif-hig}
Bands of the relative deviation of the Kurie plot in $\beta$-decay
corresponding to the allowed regions in the
$\sin^{2}2\vartheta_{ee}$--$\Delta{m}^2_{41}$
plane in Fig.~\ref{con-glo},
obtained from the GLO-HIG global analysis of short-baseline neutrino oscillation data
(see Tab~\ref{tab-bef}).
The black solid line corresponds to the best-fit point
($m_{4} = \protect1.27 \, \text{eV}$
and
$|U_{e4}|^2 = \protect0.036$).
The dashed, dotted and dash-dotted lines correspond,
respectively,
to the local minima at
($m_{4} = \protect0.95 \, \text{eV}$,
$|U_{e4}|^2 = \protect0.027$),
($m_{4} = \protect1.11 \, \text{eV}$,
$|U_{e4}|^2 = \protect0.031$) and
($m_{4} = \protect2.40 \, \text{eV}$,
$|U_{e4}|^2 = \protect0.033$).
}
\end{figure}

\begin{figure}[t!]
\begin{center}
\includegraphics*[width=\linewidth]{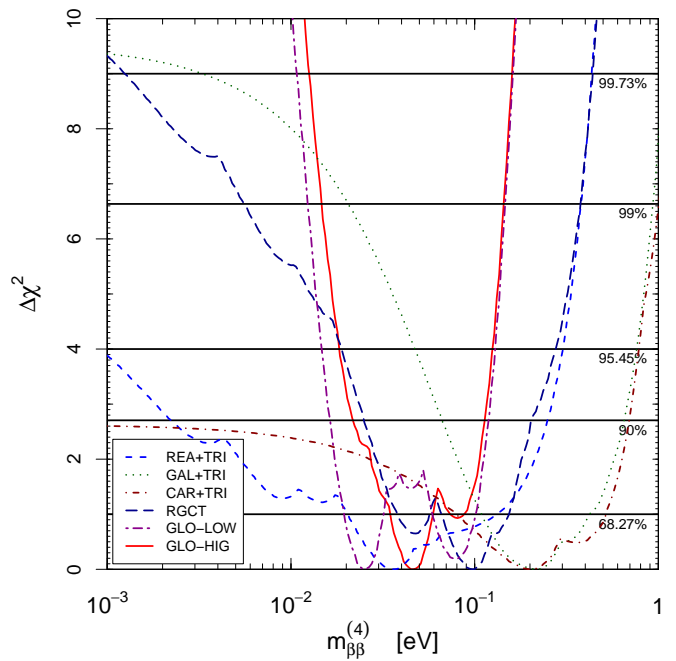}
\end{center}
\caption{ \label{mbb-chi-plt-hig}
Marginal
$\Delta\chi^2 = \chi^2 - \chi^2_{\text{min}}$
as a function of the contribution
$m_{\beta\beta}^{(4)} = |U_{e4}|^2 \sqrt{\Delta{m}^2_{41}}$
to the effective neutrinoless double-$\beta$ decay Majorana mass $m_{\beta\beta}$
obtained from
the fits of Tritium (TRI) data with
Reactor (REA),
Gallium (GAL) and
$\nu_{e}$-${}^{12}\text{C}$ (CAR)
data,
their combined fit (RGCT) and
the GLO-LOW and GLO-HIG global fits.
}
\end{figure}

\begin{figure}[t!]
\begin{center}
\includegraphics*[width=\linewidth]{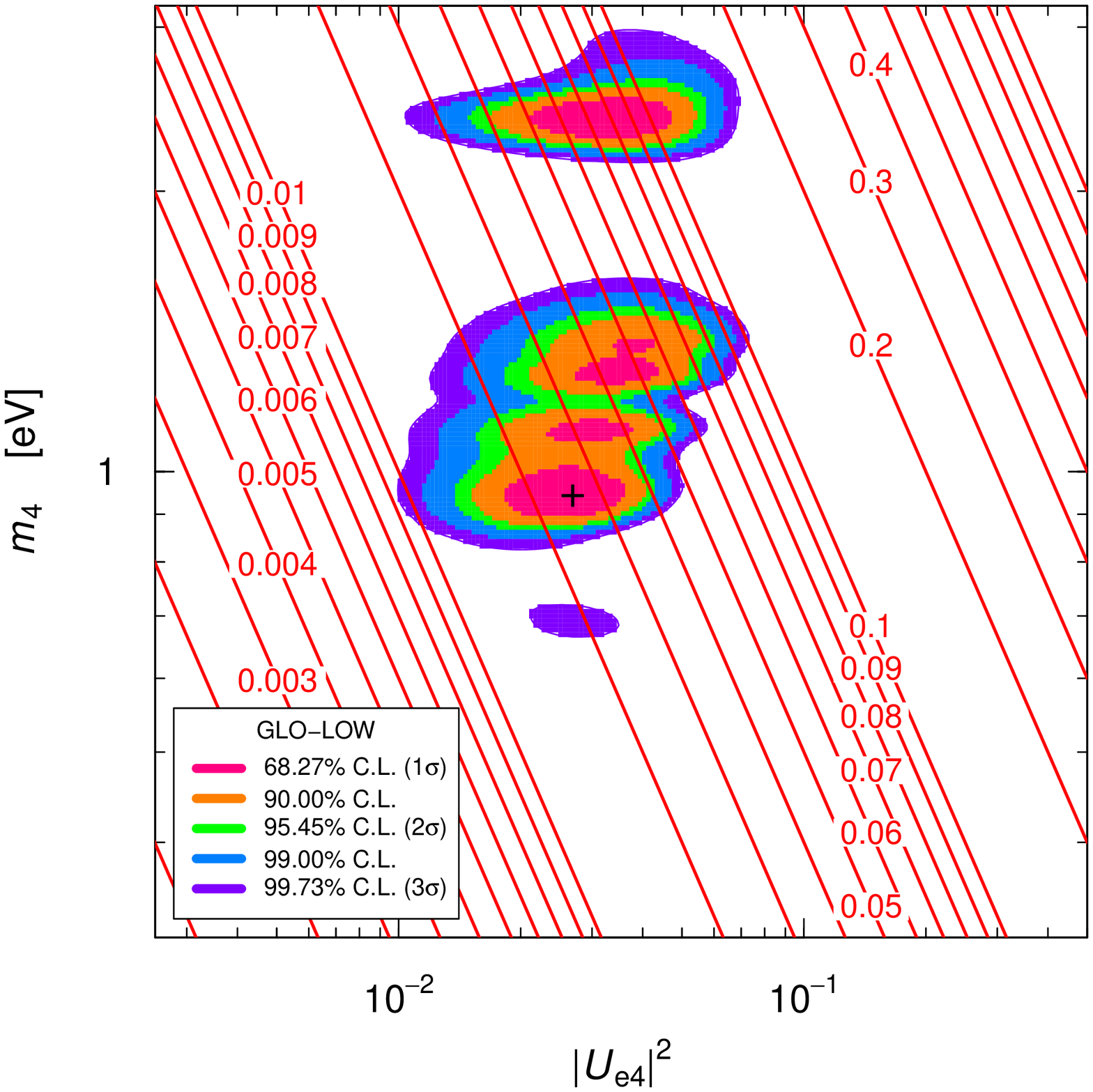}
\end{center}
\caption{ \label{con-img-see-mbb-low}
Allowed regions in the
$|U_{e4}|^2$--$m_{4}$
plane obtained
from the GLO-LOW global analysis of short-baseline neutrino oscillation data
(see Tab~\ref{tab-bef}).
The best-fit point is indicated by a cross (see Table.~\ref{tab-bef}).
The red lines have the indicated constant value of
$m_{\beta\beta}^{(4)} = |U_{e4}|^2 \sqrt{\Delta{m}^2_{41}}$.
}
\end{figure}

\begin{figure}[t!]
\begin{center}
\includegraphics*[width=\linewidth]{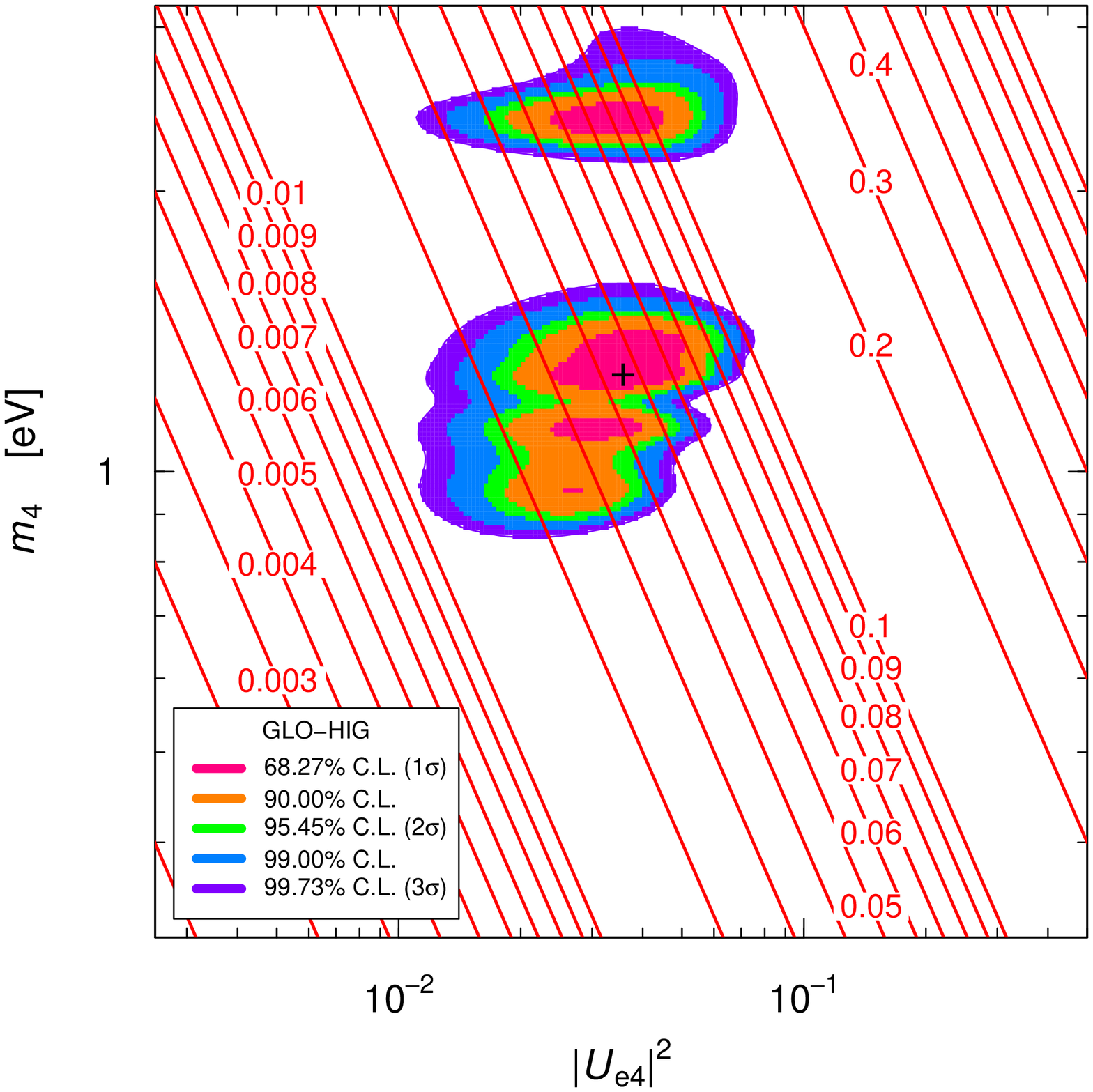}
\end{center}
\caption{ \label{con-img-see-mbb-hig}
Allowed regions in the
$|U_{e4}|^2$--$m_{4}$
plane obtained
from the GLO-HIG global analysis of short-baseline neutrino oscillation data
(see Tab~\ref{tab-bef}).
The best-fit point is indicated by a cross (see Table.~\ref{tab-bef}).
The red lines have the indicated constant value of
$m_{\beta\beta}^{(4)} = |U_{e4}|^2 \sqrt{\Delta{m}^2_{41}}$.
}
\end{figure}

\begin{table*}[t!]
\begin{center}
\setlength{\tabcolsep}{5pt}
\begin{tabular}{cccccccc}
&
&
REA
&
GAL
&
CAR
&
RGC
&
GLO-LOW
&
GLO-HIG
\\
\hline
 No Osc. &$\chi^{2}_{\text{min}}$ &27.1 &12.0 &8.2 &47.3 &195.1 &178.1 \\
 &NDF &38 &4 &10 &52 &144 &138 \\
 &GoF &0.91 &0.017 &0.61 &0.66 &0.0049 &0.019 \\
\hline 3+1 &$\chi^{2}_{\text{min}}$ &21.7 &2.3 &4.6 &36.2 &152.4 &137.5 \\
 &NDF &36 &2 &8 &50 &144 &138 \\
 &GoF &97\% &32\% &80\% &93\% &30\% &50\% \\
 &$\Delta{m}^2_{41} [\text{eV}^2]$ &1.95 &2.24 &13.80 &7.59 &0.9 &1.6 \\
 &$|U_{e4}|^2$ &0.026 &0.15 &0.13 &0.036 &0.027 &0.036 \\
 &$|U_{\mu4}|^2$ & & & & &0.021 &0.0084 \\
 &$\sin^22\vartheta_{e\mu}$ & & & & &0.0023 &0.0012 \\
 &$\sin^22\vartheta_{ee}$ &0.10 &0.51 &0.45 &0.14 &0.10 &0.14 \\
 &$\sin^22\vartheta_{\mu\mu}$ & & & & &0.083 &0.034 \\
\hline PG &$\Delta\chi^{2}_{\text{min}}$ & & & &7.6 &18.8 &11.6 \\
 &NDF & & & &4 &2 &2 \\
 &GoF & & & &11\% &0.008\% &0.3\% \\
\hline
\end{tabular}
\caption{ \label{tab-bef}
Values of
$\chi^{2}$,
number of degrees of freedom (NDF),
goodness-of-fit (GoF)
and
best-fit values
of the 3+1 oscillation parameters
obtained from
the fits of
Gallium (GAL),
Reactor (REA),
$\nu_{e}$-${}^{12}\text{C}$ (CAR)
data,
their combined fit
(RGC)
and
the global fit with (GLO-LOW) and without (GLO-HIG)
the MiniBooNE electron neutrino and antineutrino data
with reconstructed neutrino energy smaller than $475 \, \text{MeV}$.
The first three lines correspond to the case of no oscillations (No Osc.).
The following nine lines correspond to the case of 3+1 mixing.
The last three lines give the parameter goodness-of-fit (PG) \protect\cite{hep-ph/0304176}.
}
\end{center}
\end{table*}

\begin{table*}[t!]
\begin{center}
\setlength{\tabcolsep}{5pt}
\begin{tabular}{ccccccc}
&
REA+TRI
&
GAL+TRI
&
CAR+TRI
&
RGCT
&
GLO-LOW
&
GLO-HIG
\\
\hline
$m_{4}$
&
\begin{tabular}{c}
$ 1.3 - 5.2 $
\\
$ 0.41 - 13 $
\\
$ < 343 $
\end{tabular}
&
\begin{tabular}{c}
$ 1.0 - 2.8 $
\\
$ 0.85 - 5.2 $
\\
$ 0.22 - 11 $
\end{tabular}
&
\begin{tabular}{c}
$ 2.7 - 3.8 $
\\
$ < 245 $
\\
$ < 412 $
\end{tabular}
&
\begin{tabular}{c}
$ 1.4 - 3.8 $
\\
$ 1.2 - 8.4 $
\\
$ 0.42 - 16 $
\end{tabular}
&
\begin{tabular}{c}
$ 0.91 - 2.5 $
\\
$ 0.88 - 2.5 $
\\
$ 0.85 - 2.8 $
\end{tabular}
&
\begin{tabular}{c}
$ 1.2 - 2.4 $
\\
$ 0.91 - 2.5 $
\\
$ 0.87 - 2.8 $
\end{tabular}
\\
\hline
$|U_{e4}|^2$
&
\begin{tabular}{c}
$ 0.01 - 0.04 $
\\
$ 0.00 - 0.05 $
\\
$ < 0.06 $
\end{tabular}
&
\begin{tabular}{c}
$ 0.09 - 0.20 $
\\
$ 0.04 - 0.28 $
\\
$ > 0.00 $
\end{tabular}
&
\begin{tabular}{c}
$ 0.03 - 0.14 $
\\
$-$
\\
$-$
\end{tabular}
&
\begin{tabular}{c}
$ 0.02 - 0.05 $
\\
$ 0.01 - 0.06 $
\\
$ 0.00 - 0.07 $
\end{tabular}
&
\begin{tabular}{c}
$ 0.02 - 0.04 $
\\
$ 0.02 - 0.06 $
\\
$ 0.01 - 0.07 $
\end{tabular}
&
\begin{tabular}{c}
$ 0.03 - 0.05 $
\\
$ 0.02 - 0.06 $
\\
$ 0.01 - 0.07 $
\end{tabular}
\\
\hline
$m_{\beta}^{(4)}$
&
\begin{tabular}{c}
$ 0.16 - 0.81 $
\\
$ 0.02 - 1.69 $
\\
$ < 2.29 $
\end{tabular}
&
\begin{tabular}{c}
$ 0.36 - 1.06 $
\\
$ 0.23 - 1.79 $
\\
$ 0.05 - 2.29 $
\end{tabular}
&
\begin{tabular}{c}
$ 0.48 - 1.40 $
\\
$ < 1.79 $
\\
$ < 2.29 $
\end{tabular}
&
\begin{tabular}{c}
$ 0.23 - 0.74 $
\\
$ 0.15 - 1.45 $
\\
$ 0.02 - 2.15 $
\end{tabular}
&
\begin{tabular}{c}
$ 0.14 - 0.49 $
\\
$ 0.12 - 0.56 $
\\
$ 0.10 - 0.65 $
\end{tabular}
&
\begin{tabular}{c}
$ 0.21 - 0.45 $
\\
$ 0.13 - 0.56 $
\\
$ 0.11 - 0.63 $
\end{tabular}
\\
\hline
$m_{\beta\beta}^{(4)}$
&
\begin{tabular}{c}
$ 0.021 - 0.13 $
\\
$ < 0.30 $
\\
$ < 0.42 $
\end{tabular}
&
\begin{tabular}{c}
$ 0.110 - 0.41 $
\\
$ 0.048 - 0.76 $
\\
$ 0.003 - 1.03 $
\end{tabular}
&
\begin{tabular}{c}
$ 0.081 - 0.51 $
\\
$ < 0.78 $
\\
$ < 1.17 $
\end{tabular}
&
\begin{tabular}{c}
$ 0.039 - 0.15 $
\\
$ 0.019 - 0.28 $
\\
$ 0.001 - 0.44 $
\end{tabular}
&
\begin{tabular}{c}
$ 0.020 - 0.10 $
\\
$ 0.015 - 0.13 $
\\
$ 0.011 - 0.16 $
\end{tabular}
&
\begin{tabular}{c}
$ 0.035 - 0.09 $
\\
$ 0.018 - 0.12 $
\\
$ 0.013 - 0.16 $
\end{tabular}
\\
\hline
\end{tabular}
\caption{ \label{tab-rng}
Allowed
$1\sigma$,
$2\sigma$ and
$3\sigma$
ranges of
$m_{4}=\sqrt{\Delta{m}^2_{41}}$,
$|U_{e4}|^2$,
$m_{\beta}^{(4)}=|U_{e4}|\sqrt{\Delta{m}^2_{41}}$
and
$m_{\beta\beta}^{(4)}=|U_{e4}|^2\sqrt{\Delta{m}^2_{41}}$
obtained from the fits of Tritium data
with
Gallium (GAL+TRI),
Reactor (REA+TRI) and
$\nu_{e}$-${}^{12}\text{C}$ (CAR+TRI)
data,
their combined fit (RGCT) and
the global fit with (GLO-LOW) and without (GLO-HIG)
the MiniBooNE electron neutrino and antineutrino data
with reconstructed neutrino energy smaller than $475 \, \text{MeV}$.
Masses are given in eV.
}
\end{center}
\end{table*}

The global fit presented in this paper upgrades that presented in
Ref.~\cite{1109.4033}
by the addition of
$\nu_{e} + {}^{12}\text{C} \to {}^{12}\text{N}_{\text{g.s.}} + e^{-}$
scattering data, as suggested in Ref.~\cite{1106.5552}.
The considered sets of data are:

\begin{itemize}

\item
The short-baseline
$\boss{\nu}{\mu}\to\boss{\nu}{e}$
data
of the
LSND \cite{hep-ex/0104049},
KARMEN \cite{hep-ex/0203021},
NOMAD \cite{hep-ex/0306037}
and 
MiniBooNE neutrino \cite{0812.2243} and antineutrino \cite{1007.1150,Zimmerman-PANIC2011,Djurcic-NUFACT2011}
experiments.

\item
The short-baseline $\bar\nu_{e}$ disappearance data
of the
Bugey-3 \cite{Declais:1995su},
Bugey-4 \cite{Declais:1994ma},
ROVNO91 \cite{Kuvshinnikov:1990ry},
Gosgen \cite{Zacek:1986cu},
ILL \cite{Hoummada:1995zz}
and
Krasnoyarsk \cite{Vidyakin:1990iz}
reactor antineutrino experiments,
taking into account the new calculation of the reactor $\bar\nu_{e}$ flux
\cite{1101.2663,1106.0687}
which indicates a small $\bar\nu_{e}$ disappearance
(the reactor antineutrino anomaly \cite{1101.2755}),
and the KamLAND \cite{0801.4589} bound on $|U_{e4}|^2$
(see Ref.~\cite{1012.0267}).

\item
The short-baseline $\nu_{\mu}$ disappearance data
of the CDHSW experiment
\cite{Dydak:1984zq},
the constraints on $|U_{\mu4}|^2$ obtained in Ref.~\cite{0705.0107}
from the analysis of
the data of
atmospheric neutrino oscillation experiments,
and
the bound on $|U_{\mu4}|^2$ obtained from MINOS neutral-current data \cite{1104.3922}
(see Refs.~\cite{1105.5946,1109.4033}).

\item
The
data of Gallium radioactive source experiments
(GALLEX
\cite{Anselmann:1995ar,Hampel:1998fc,1001.2731}
and
SAGE
\cite{Abdurashitov:1996dp,hep-ph/9803418,nucl-ex/0512041,0901.2200})
which indicate a $\nu_{e}$ disappearance
(the Gallium neutrino anomaly
\cite{hep-ph/9411414,Laveder:2007zz,hep-ph/0610352,0707.4593,0711.4222,0902.1992,1005.4599,1006.2103,1006.3244,1101.2755}).
We analyze the Gallium data according to Ref.~\cite{1006.3244}.

\item
The
$\nu_{e} + {}^{12}\text{C} \to {}^{12}\text{N}_{\text{g.s.}} + e^{-}$
scattering data of the
KARMEN \cite{Bodmann:1994py,hep-ex/9801007}
and
LSND \cite{hep-ex/0105068}
experiments,
which constrain the short-baseline $\nu_{e}$ disappearance
\cite{1106.5552}.

\end{itemize}

The plan of the paper is as follows.
Since this is the first time that
the KARMEN and LSND $\nu_{e} + {}^{12}\text{C} \to {}^{12}\text{N}_{\text{g.s.}} + e^{-}$
scattering data are taken into account in a global fit of neutrino oscillation short-baseline data,
in Section~\ref{glo}
we discuss the method of the analysis and
we present the ensuing upgrade of the global fit published in Ref.~\cite{1109.4033}.
In Section~\ref{beta}
we discuss the predictions for the effective electron neutrino mass in $\beta$-decay.
In Section~\ref{Neutrinoless}
we present the predictions for the effective Majorana neutrino mass in neutrinoless double-$\beta$-decay.
In Section~\ref{Conclusions}
we draw our conclusions.

\section{Global Fit}
\label{glo}

In this section we present the upgrade of the results
of the global fit of short-baseline neutrino oscillation data
obtained in Ref.~\cite{1109.4033}
with the inclusion of the
KARMEN \cite{Bodmann:1994py,hep-ex/9801007}
and
LSND \cite{hep-ex/0105068}
$\nu_{e} + {}^{12}\text{C} \to {}^{12}\text{N}_{\text{g.s.}} + e^{-}$
scattering data suggested in Ref.~\cite{1106.5552}.

We analyzed the KARMEN and LSND data summarized in Ref.~\cite{1106.5552}
without assuming a theoretical model for the
$\nu_{e} + {}^{12}\text{C} \to {}^{12}\text{N}_{\text{g.s.}} + e^{-}$
cross section.
We assumed
only a dependence of the cross section on
$(E-Q)^2$,
where $Q=17.3\,\text{MeV}$
is the $Q$-value of the reaction.
Such dependence on the neutrino energy $E$
is due to the allowed character of the transition from
the $0^+$ ground state of ${}^{12}\text{C}$
to
the $1^+$ ground state of ${}^{12}\text{N}$.
The information on neutrino oscillations comes from the different source-detector distances in KARMEN and LSND:
$L_{\text{KARMEN}} = 17.7 \, \text{m}$
and
$L_{\text{LSND}} = 29.8 \, \text{m}$.
The best fit values of the oscillation parameters
and
the 95\% C.L. allowed region in the
$\sin^{2}2\vartheta_{ee}$--$\Delta{m}^{2}_{41}$ plane
are given, respectively, in Tab.~\ref{tab-bef} and Fig.~\ref{sup-gal-rea-nec},
together with the corresponding quantities obtained from the analysis of
reactor and Gallium data.

The $\nu_{e}-{}^{12}\text{C}$ curve in Fig.~\ref{sup-gal-rea-nec} is an exclusion curve
which proscribes the region of the oscillation parameters on the right.
The best-fit is obtained for rather large values of the oscillation parameters:
$\Delta{m}^{2}_{41}=13.80\,\text{eV}^2$
and
$\sin^{2}2\vartheta_{ee}=0.45$.
The reason of the improvement of the value of $\chi^{2}_{\text{min}}$
obtained with the best-fit values of the oscillation parameters
with respect to the case of no oscillations can be understood from Fig.~\ref{hst-nec},
where one can see that there is an improvement in the fit of the
KARMEN data due to the shorter source-detector distance with respect to the LSND data,
for which there is practically no improvement,
because oscillations are averaged out.
From Fig.~\ref{sup-gal-rea-nec}
one can see that for $\Delta{m}^{2}_{41} \gtrsim 20\,\text{eV}^2$
there is no constraint on neutrino oscillations
from $\nu_{e}-{}^{12}\text{C}$
because the oscillations are averaged out in both the
KARMEN and LSND experiments.

The comparison of the 95\% C.L.
reactor, Gallium and $\nu_{e}-{}^{12}\text{C}$ allowed regions in the
$\sin^{2}2\vartheta_{ee}$--$\Delta{m}^{2}_{41}$ plane
presented in Fig.~\ref{sup-gal-rea-nec}
shows that there is a region of overlap for
$0.1 \lesssim \sin^{2}2\vartheta_{ee} \lesssim 0.2$
and
$\Delta{m}^{2}_{41} \gtrsim 2\,\text{eV}^2$,
indicating that the results of the three sets of data are compatible
with the hypothesis of neutrino oscillations.

Figure~\ref{sup-gal-rea-nec}
shows also the 95\% C.L. allowed region in the
$\sin^{2}2\vartheta_{ee}$--$\Delta{m}^{2}_{41}$ plane
obtained from the combined analysis of reactor, Gallium and $\nu_{e}-{}^{12}\text{C}$,
with the best-fit values of the oscillation parameters listed in
Tab.~\ref{tab-bef}.
The compatibility of the three sets of data
is confirmed by the
11\%
value of the parameter goodness-of-fit
\cite{hep-ph/0304176}.

Figure~\ref{con-glo} shows the allowed regions in the
$\sin^{2}2\vartheta_{e\mu}$--$\Delta{m}^2_{41}$,
$\sin^{2}2\vartheta_{ee}$--$\Delta{m}^2_{41}$ and
$\sin^{2}2\vartheta_{\mu\mu}$--$\Delta{m}^2_{41}$
planes obtained from the global analysis of short-baseline neutrino oscillation data
listed in the Introduction.
We made two global analyses named
GLO-LOW and GLO-HIG, respectively,
with and without
the three MiniBooNE electron neutrino and antineutrino bins
with reconstructed neutrino energy smaller than $475 \, \text{MeV}$,
which have an excess of events called the "MiniBooNE low-energy anomaly".
The best-fit values of the oscillation parameters are listed in
Tab.~\ref{tab-bef}.
In the global analyses we consider values of
$\Delta{m}^2_{41}$
smaller than $10\,\text{eV}^2$,
since larger values are strongly incompatible with
the cosmological constraints on neutrino masses
\cite{1006.5276,1102.4774,1104.0704,1104.2333,1106.5052,1108.4136,1109.2767,1110.4271}.

The GLO-LOW and GLO-HIG
analyses are similar,
respectively,
to the LOW-GAL and HIG-GAL analyses presented in Ref.~\cite{1109.4033},
but take into account in addition the KARMEN and LSND $\nu_{e}-{}^{12}\text{C}$
scattering data.
From Fig.~\ref{con-glo} and Tab.~\ref{tab-bef}
one can see that the contribution of the $\nu_{e}-{}^{12}\text{C}$ data
is beneficial for the lowering of the best-fit value of
$\Delta{m}^2_{41}$
from $5.6\,\text{eV}^2$ obtained in Ref.~\cite{1109.4033}
to
$0.9\,\text{eV}^2$ in GLO-LOW
and
$1.6\,\text{eV}^2$ in GLO-HIG.
These values of $\Delta{m}^2_{41}$
are more compatible with the cosmological constraints on neutrino masses
\cite{1006.5276,1102.4774,1104.0704,1104.2333,1106.5052,1108.4136,1109.2767,1110.4271}.

Comparing the GLO-LOW and GLO-HIG parts of Fig.~\ref{con-glo} and Tab.~\ref{tab-bef}
one can see that the inclusion of the fit of
the MiniBooNE low-energy data
favors small values of
$\Delta{m}^2_{41}$.
This fact has been noted and explained in Ref.~\cite{1109.4033}.
Hence,
the results of the GLO-LOW analysis are more attractive than those of the GLO-HIG in view of
a better compatibility with
cosmological constraints on the neutrino masses.

The well known tension between appearance and disappearance data
\cite{hep-ph/9607372,hep-ph/9903454,hep-ph/0207157,hep-ph/0305255,hep-ph/0405172,hep-ph/0609177,0705.0107,1007.4171,0906.1997,1007.4171,1012.0267,1103.4570,1107.1452,1109.4033,1110.3735}
is slightly worsened by the inclusion in the fit of
the $\nu_{e}-{}^{12}\text{C}$ scattering data,
which strengthen the disappearance constrain.

In the GLO-LOW analysis the
0.008\%
parameter goodness-of-fit,
has worsened with respect to the
0.04\% obtained in Ref.~\cite{1109.4033},
which was already so low that
we wrote that the MiniBooNE low-energy anomaly
"may have an explanation different from
$\boss{\nu}{\mu}\to\boss{\nu}{e}$
oscillations".
Nevertheless,
in the following we will discuss
the GLO-LOW predictions on the effective neutrino masses measured in
$\beta$-decay
and
neutrinoless double-$\beta$-decay
experiments,
because we are not aware of an a-priori argument which allows to
exclude from the analysis
the MiniBooNE low-energy data.
The exclusion a-posteriori motivated by the results of the fit
may be hazardous,
taking also into account the nice value of the global
goodness-of-fit
(30\%)
and the above-mentioned preference for
small values of $\Delta{m}^2_{41}$ in agreement with the same preference of the cosmological data.

Considering the GLO-HIG analysis,
the
0.3\%
appearance-disappearance
parameter goodness-of-fit is
lower than that obtained in Ref.~\cite{1109.4033}
(1\%)
because of the above-mentioned
strengthening of the disappearance constraint induced by the
$\nu_{e}-{}^{12}\text{C}$ scattering data.
However,
the appearance-disappearance
parameter goodness-of-fit
is not dramatically low and the fit cannot be rejected,
also taking into account the pleasant
50\%
value of the global goodness-of-fit.

In the next two sections we discuss the predictions for the effective neutrino masses measured in
$\beta$-decay
and
neutrinoless double-$\beta$-decay
experiments.

\section{$\beta$-Decay}
\label{beta}

The effective electron neutrino mass $m_{\beta}$ in $\beta$-decay experiments is given by
\cite{Shrock:1980vy,McKellar:1980cn,Kobzarev:1980nk,hep-ph/0012018}
(other approaches are discussed in Refs.~\cite{hep-ph/0105105,hep-ph/0110232,hep-ph/0211341})
\begin{equation}
m_{\beta}^2
=
\sum_{k} |U_{ek}|^2 m_{k}^2
\,.
\label{029}
\end{equation}
The most accurate measurements of $m_{\beta}$
have been obtained in the
Mainz \cite{hep-ex/0412056}
and
Troitsk \cite{Lobashev:2003kt}
experiments,
whose combined upper bound is
\cite{1005.4599}
\begin{equation}
m_{\beta} \leq 1.8 \, \text{eV}
\qquad
(\text{Mainz+Troitsk, 95\% C.L.})
\,.
\label{034}
\end{equation}

In the hierarchical 3+1 scheme
we have the lower bound
\begin{equation}
m_{\beta}
\geq
|U_{e4}| \sqrt{\Delta{m}^2_{41}}
\equiv
m_{\beta}^{(4)}
\,.
\label{mb4}
\end{equation}
Therefore,
from the analysis of short-baseline neutrino oscillation data we
can derive predictions for the possibility of observing a neutrino mass effect
in the KATRIN experiment \cite{1110.0087},
which is under construction and scheduled to start in 2012,
and in other possible future experiments.

Let us however note that the effective electron neutrino mass
in Eq.~(\ref{029}) has been derived assuming that all the neutrino masses are smaller
than the experimental energy resolution
(see Ref.~\cite{Giunti-Kim-2007}).
If $m_{4}$ is of the order of 1 eV,
the approximation is acceptable for the interpretation of the result of the
Mainz and Troitsk
experiments, which had, respectively,
energy resolutions of
4.8 eV and 3.5 eV \cite{0912.1619}.
On the other hand,
the energy resolution of the KATRIN experiment will be
0.93 eV near the end-point of the energy spectrum of the electron emitted in Tritium decay,
at
$T=Q$,
where $T$ is the kinetic energy of the electron
and
$Q=18.574\,\text{keV}$
is the $Q$-value of the decay.
If the value of $m_{4}$ is larger than the energy resolution of the experiment,
its effect on the measured electron spectrum
cannot be summarized by one effective quantity,
because the Kurie function $K(T)$ is given by
\begin{align}
\null & \null
\frac{K^2(T)}{Q-T}
=
\sqrt{(Q-T)^2-\widetilde{m}_{\beta}^2}
-
|U_{e4}|^2
(Q-T)
\nonumber
\\
\null & \null
+
|U_{e4}|^2
\sqrt{(Q-T)^2-m_{4}^{2}}
\
\theta(Q-T-m_{4})
\,,
\label{kurie}
\end{align}
where
$
\widetilde{m}_{\beta}^2
=
\sum_{k=1}^{3} |U_{ek}|^2 m_{k}^2
$
is the contribution of the three neutrino masses much smaller than 1 eV
and $\theta$ is the Heaviside step function.

In the following we discuss
the predictions obtained from the
GLO-LOW and GLO-HIG analyses
of short-baseline neutrino oscillation data
for both the contribution
$m_{\beta}^{(4)}$
to the effective mass in $\beta$-decay
and the
distortion
of the Kurie function due to $m_{4}$.

Figure~\ref{mbt-chi-plt}
shows the marginal
$\Delta\chi^2 = \chi^2 - \chi^2_{\text{min}}$
as a function of $m_{\beta}^{(4)}$
for the data analyses listed in Tab.~\ref{tab-rng},
which gives the
$1\sigma$,
$2\sigma$ and
$3\sigma$
allowed ranges of
$m_{4}$,
$|U_{e4}|^2$
and
$m_{\beta}^{(4)}$.

The predictions for
$m_{\beta}^{(4)}$
obtained from the
analyses of
reactor, Gallium and $\nu_{e}-{}^{12}\text{C}$ data
are interesting, because
these data give direct information on
$|U_{e4}|$
from the amplitude
of
$\boss{\nu}{e}$
disappearance
(see Eq.~(\ref{survisin})).
Since the data of these experiments do not allow us to constrain the upper value of
$\Delta{m}^2_{41}$,
as can be seen in Fig.~\ref{sup-gal-rea-nec},
we have analyzed their data in combination with the Tritium data
of
the Mainz \cite{hep-ex/0412056}
and
Troitsk \cite{Lobashev:2003kt}
experiment.
This is the origin of the corresponding upper limits
for $m_{4}$ and $m_{\beta}^{(4)}$ in Tab.~\ref{tab-rng}
and
the corresponding steep rise of
$\Delta\chi^2$
for
$m_{\beta}^{(4)} \gtrsim 2 \, \text{eV}$
in Fig.~\ref{mbt-chi-plt}.

As one can see from
Fig.~\ref{mbt-chi-plt} and Tab.~\ref{tab-rng},
the most stringent predicted ranges for $m_{\beta}^{(4)}$
are obtained in the GLO-LOW and GLO-HIG analyses,
but all the analyses agree in favoring values of $m_{\beta}^{(4)}$
between about 0.1 and 0.7 eV,
which is promising for the perspectives of future experiments.

In Figures~\ref{plt-kur-dif-low} and \ref{plt-kur-dif-hig}
we show the relative deviation of the Kurie function
from the massless case
($K(T) = Q - T$)
obtained in the GLO-LOW and GLO-HIG analyses,
neglecting the contribution of
$\widetilde{m}_{\beta}$
in Eq.~(\ref{kurie}).
For $T > Q - m_{4}$
the deviation is constant, because the Kurie function in Eq.~(\ref{kurie})
reduces to
\begin{equation}
K(T)
=
(Q-T) \sqrt{1 - |U_{e4}|^2}
\,.
\label{kurie-1}
\end{equation}
For
$T = Q - m_{4}$
there is a kink and
for $T < Q - m_{4}$
the Kurie function depends on both
$m_{4}$ and $|U_{e4}|^2$,
as given by Eq.~(\ref{kurie}).

From Figs.~\ref{plt-kur-dif-low} and \ref{plt-kur-dif-hig}
one can see that high precision will be needed in order to see the effect of $m_{4}$
and measure $|U_{e4}|^2$,
which is the only parameter which determines the deviation of $K(T)$ from the massless Kurie function near the end point,
for $T > Q - m_{4}$.
If the mixing parameters are near the best-fit point of the GLO-LOW analysis,
a precision of about one percent will be needed within 1 eV from the end-point of the spectrum.
Finding the effect of $m_{4}$ farther from the end-point,
for $T < Q - m_{4}$
is more difficult,
because the relative deviation of the Kurie function can be as small as about $10^{-3}$.
The GLO-HIG analysis prefers slightly larger values of $m_{4}$,
but the discovery of an effect in $\beta$-decay
will require a similar precision.

\section{Neutrinoless $\beta\beta$-Decay}
\label{Neutrinoless}

If massive neutrinos are Majorana particles,
neutrinoless double-$\beta$ decay is possible,
with a decay rate proportional to the effective Majorana mass
(see Refs.~\cite{hep-ph/0202264,hep-ph/0211462,hep-ph/0405078,0708.1033,Giunti-Kim-2007,1106.1334})
\begin{equation}
m_{\beta\beta}
=
\left| \sum_{k} U_{ek}^2 m_{k} \right|
\,.
\label{050}
\end{equation}
The results of the analysis of short-baseline oscillation data
allow us to calculate the contribution of the heaviest massive neutrino $\nu_{4}$
to $m_{\beta\beta}$, which is given by
\begin{equation}
m_{\beta\beta}^{(4)}
=
|U_{e4}|^2 \sqrt{\Delta{m}^2_{41}}
\,,
\label{m2b4}
\end{equation}
taking into account the mass hierarchy in Eq.~(\ref{hierarchy}).
If there are no unlikely cancellations among the contributions of
$m_{1}$,
$m_{2}$,
$m_{3}$
and that of
$m_{4}$
\cite{hep-ph/9906275}
(possible cancellations are discussed in Refs.\cite{hep-ph/0512234,1110.5795}),
the value of
$m_{\beta\beta}^{(4)}$
is a lower bound for the effective neutrino mass
which could be observed
in future
neutrinoless double-$\beta$ decay experiments
(see the review in Ref.~\cite{1109.5515}).

Figure~\ref{mbb-chi-plt-hig}
shows the marginal
$\Delta\chi^2$
as a function of
$m_{\beta\beta}^{(4)}$
for the data analyses listed in Tab.~\ref{tab-rng},
which gives the
$1\sigma$,
$2\sigma$ and
$3\sigma$
allowed ranges of $m_{\beta\beta}^{(4)}$.

The analyses of the
Gallium and
$\nu_{e}$-${}^{12}\text{C}$ scattering data
give optimistic indications
in favor of a value of $m_{\beta\beta}^{(4)}$
of about 0.2 eV,
albeit with large uncertainties,
but their combined fit with reactor data
lowers the prediction
to a best-fit value of about 0.1 eV,
with a $1\sigma$ allowed range which extends down to
about 0.04 eV.
The reason is that reactor data constrain $|U_{e4}|^2$
to small values, as can be seen in Tab.~\ref{tab-rng}.

The predictions for $m_{\beta\beta}^{(4)}$
obtained from global GLO-LOW and GLO-HIG agree in indicating a $3\sigma$
allowed range between about 0.01 and 0.1 eV.
The connection of the value of $m_{\beta\beta}^{(4)}$
with the allowed regions for the oscillation parameters
is clarified in Figs.~\ref{con-img-see-mbb-low} and \ref{con-img-see-mbb-hig},
where we show the allowed regions in the
$|U_{e4}|^2$--$m_{4}$
plane obtained, respectively,
from the GLO-LOW and GLOW-HIG analyses,
together with lines of constant $m_{\beta\beta}^{(4)}$.
One can see that if the oscillation parameters are close to the best-fit point of the GLO-LOW analysis,
at
$m_{4} = 0.94 \, \text{eV}$,
which is favored by cosmological data,
the value of $m_{\beta\beta}^{(4)}$
is about 0.02-0.03 eV.
In order to have a large value of $m_{\beta\beta}^{(4)}$,
around 0.1 eV,
the oscillation parameters must lie in the large-$m_{4}$ region at
$m_{4} \simeq 2.40 \, \text{eV}$,
or on the large-$|U_{e4}|^2$ border
of the allowed region at
$m_{4} \simeq 1.27 \, \text{eV}$.

\section{Conclusions}
\label{Conclusions}

In this paper we presented an upgrade
of the
3+1 global fit of short-baseline neutrino oscillation data presented in Ref.~\cite{1109.4033}
obtained with the addition of
KARMEN \cite{Bodmann:1994py,hep-ex/9801007}
and
LSND \cite{hep-ex/0105068}
$\nu_{e} + {}^{12}\text{C} \to {}^{12}\text{N}_{\text{g.s.}} + e^{-}$
scattering data, as suggested in Ref.~\cite{1106.5552}.
We have shown that the new data favor low values of
$\Delta{m}^2_{\text{41}}$,
which are appealing in view of the cosmological constraints on neutrino masses
\cite{1006.5276,1102.4774,1104.0704,1104.2333,1106.5052,1108.4136,1109.2767,1110.4271}.

We discussed the implications for the measurements of the effective neutrino mass
in $\beta$-decay and neutrinoless double-$\beta$-decay
experiments.

The predicted contribution of $m_{4}$ to the effective neutrino mass $m_{\beta}$
in $\beta$-decay is in the range between about 0.1 and 0.7 eV,
most of which will be explored after 2012 by
the KATRIN experiment \cite{1110.0087},
which will have a sensitivity of about 0.2 eV.
If KATRIN or future experiments
will have a precision of about one percent near the end-point of the spectrum,
for $T > Q - m_{4}$,
the effects of
$m_{4}$ and $|U_{e4}|^2$ may be measured separately.

The predicted contribution of $m_{4}$ to the effective Majorana mass $m_{\beta\beta}$
in neutrinoless double-$\beta$ decay
is in the range between about 0.01 and 0.1 eV.
The expected value of $m_{\beta\beta}$ is smaller than the expected value of $m_{\beta}$
because it is suppressed by an additional power of $|U_{e4}|$, which is small.
Nevertheless,
it is important that
$m_{\beta\beta}$
has a lower limit
(if no unlikely cancellations among the four mass contributions occur \cite{hep-ph/9906275})
which is an important motivation for the development of neutrinoless double-$\beta$ decay experiments
\cite{1109.5515}.

\bigskip
\centerline{\textbf{Acknowledgments}}
\medskip

We would like to thank
W.C. Louis
for interesting discussions.

\bibliography{bibtex/nu}

\end{document}